\documentclass[prl,showpacs,superscriptaddress,twocolumn]{revtex4-2}

\usepackage[makeroom]{cancel}

\usepackage{amsmath}    
\usepackage{amssymb}
\usepackage{graphicx}   
\usepackage{verbatim}   
\usepackage{color}      
\usepackage{hyperref}   
\usepackage[normalem]{ulem}
\usepackage{natbib}
\usepackage{enumitem}
\usepackage{empheq}

\hypersetup{colorlinks,linkcolor=blue,urlcolor=blue,citecolor=blue}

\definecolor{darkgreen}{rgb}{0,0.5,0}
\definecolor{purple}{rgb}{0.35,0,0.35}
\definecolor{orange}{rgb}{1,0.5,0}
\definecolor{darkred}{rgb}{.7,0,0}
\definecolor{darkblue}{rgb}{0,0,.3}
\definecolor{grey}{rgb}{.6,.6,.6}
\definecolor{dimgreen}{rgb}{0.2,0.6,0.1}

\definecolor{darkgreen}{rgb}{0,0.5,0}

\begin{document}

\newcommand{\jav}[1]{{\color{red}#1}}

\title{Momentum space magic for the transverse field quantum Ising model}

\author{Bal\'azs D\'ora}
\email{dora.balazs@ttk.bme.hu}
\affiliation{Department of Theoretical Physics, Institute of Physics, Budapest University of Technology and Economics, M\H uegyetem rkp. 3., H-1111
Budapest, Hungary}
\author{C\u{a}t\u{a}lin Pa\c{s}cu Moca}
\affiliation{MTA-BME Quantum Dynamics and Correlations Research Group, Institute of Physics, Budapest University of Technology and Economics, M\H uegyetem rkp. 3., H-1111, Budapest, Hungary}
\affiliation{Department  of  Physics,  University  of  Oradea,  410087,  Oradea,  Romania}

\date{\today}

\begin{abstract}
Stabilizer entropies and quantum magic have been extensively explored in real-space formulations of quantum systems within the framework of resource theory. 
However, interesting and transparent physics often emerges in momentum space, such as Cooper pairing.
Motivated by this, we investigate the \emph{momentum-space} structure of Pauli strings and stabilizer entropies in the one-dimensional 
transverse-field quantum Ising model. By mapping the Ising chain onto momentum-space qubits, where the stabilizer state corresponds to the 
paramagnetic state, we analyze the evolution of the Pauli string distribution. In the ferromagnetic phase, the distribution is broad, whereas in 
the paramagnetic phase, it develops a two-peaked structure. 
We demonstrate that all ferromagnetic states possess the same degree of magic in the thermodynamic limit, while stabilizer entropies are 
non-analytic  at the critical point and vanish with increasing transverse field.
 The momentum-space approach to quantum magic not only 
complements its real-space counterpart but also provides advantages in terms of analyzing nonstabilizerness and classical simulability. 

\end{abstract}

\maketitle

\paragraph{Introduction.}
Quantum computation and information theory\cite{nielsen,vedral} holds the promise to analyze and solve problems, which are beyond the reach of capabilities of classical computers. 
One relevant problem constitutes the simulation
of quantum many-body systems\cite{smith2019} for the purpose of gaining information about a particular model or using it to perform a given computational task\cite{childsrmp}.
The basic idea behind quantum information science relies on entanglement\cite{horodecki,amico,eisert,preskill} in achieving this goal. However, there exist highly entangled states which can still be simulated efficiently classically. 
These stabilizer states are built from 
Clifford circuits\cite{gottesman,aaronson,gidney}, that are generated using the Hadamard, phase or 
controlled-not gates. The amount of deviations of a quantum state from stabilizer states is quantified by nonstabilizerness or magic\cite{bravyi,Veitch_2014,leone2022,tarabunga,liuprxq,oliviero2022,lami,haug,TarabungaQuantum}.

Oftentimes, quantum simulations are based on real space local Hamiltonians. However, momentum-space Hamiltonians can be more illuminating compared to the real space version 
and  reveal the origin of  various instabilities and phase transitions.
For example, Cooper pairs are made of particles with opposite momentum and give rise to superconductivity\cite{tinkham}. 
In one dimension,  Luttinger liquids physics appear after coupling right- and left-moving fermions together  \cite{giamarchi,nersesyan}.
Several quantum spin models admit a particularly simple form or even exact solutions in momentum space, such as the transverse field 
Ising chain\cite{sachdev} or Kitaev's spin liquid model on the honeycomb
lattice\cite{kitaev2006}. Simulating various aspects of the transverse field quantum Ising model in momentum space represents a fruitful 
enterprise\cite{cui2020,gong2016}, and many theoretical ideas have been tested 
experimentally in this manner for closed and non-hermitian\cite{PRXQ} quantum systems.

Therefore, it looks natural to extend the ideas of quantum simulations to momentum space and ask how challenging it is to realize and simulate these 
states on quantum hardware. In order to answer this question, we need
to map our model of interest to momentum space qubits, determine the stabilizer states and investigate the measures of nonstabilizerness. 
In the present work, we achieve this for the transverse-field quantum Ising chain. By performing a Jordan-Wigner transformation followed by a 
Fourier transformation, and then applying an inverse Jordan-Wigner transformation, we map the model onto qubits residing in momentum space. We 
then analyze the distribution of Pauli strings, as well as the magic gap and stabilizer entropies.


Most importantly, momentum space magic can be evaluated for arbitrary system sizes and reveals the presence of a quantum critical point.
We observe that momentum-space simulations offer advantages over their real-space counterparts, as the magic can be parametrically smaller in momentum space.
The momentum-space basis has been explored in the context of quantum computation for lattice gauge theories and nuclear two-body systems\cite{muellerprxq,weiss24,Thompson_2022}.

\paragraph{From real to momentum space Ising model.}
The one dimensional ferromagnetic transverse field Ising model for $N$ spins (assumed to be even) is given by
\begin{gather}
H=-J\sum_{n=1}^N\left(\sigma^x_n\sigma^x_{n+1}+g\sigma^z_n\right),
\end{gather}
where $J>0$ represents the unit of energy, $g$ is the dimensionless transverse field and periodic boundary conditions are used. The Pauli matrices act on the lattice site $n$.
The model undergoes a quantum phase transition\cite{sachdev} from a ferromagnetic phase to a paramagnetic phase at the quantum critical point $|g|=1$.
After a Jordan-Wigner transformation\cite{dziarmaga,duttaising,Calabrese_2012}, the model is mapped onto spinless fermions as
\begin{gather}
H=J\sum_k2(g-\cos(k))c^+_kc_k+\nonumber\\
+i\sin(k)\left(c^+_kc^+_{-k}+c_kc_{-k}\right)-g,
\label{hamk}
\end{gather}
and the lattice constant is taken to be unity. The wavenumber is half integer quantized as $k=\frac{2\pi}{N}(-\frac{N-1}{2}:1:\frac{N-1}{2})$ and the quasiparticle 
spectrum is $E(k)=2J\sqrt{g^2-2g\cos(k)+1}$.
The ground state wavefunction of this Hamiltonian is of BCS form\cite{tinkham} as
\begin{gather}
|\Psi\rangle=\bigotimes_{k>0}|\Psi_k\rangle,\hspace*{4mm} |\Psi_k\rangle =\left(u_k+iv_kc^+_kc^+_{-k}\right)|vac\rangle_{k, -k},
\label{wf}
\end{gather}
where $|vac\rangle_{k,-k}$ is the fermionic vacuum state for $k$ and $-k$ modes and $(u_k,v_k)=(\cos(\theta_k/2),\sin(\theta_k/2))$ with $\tan(\theta_k)=\sin(k)/(g-\cos(k))$.
The true ground state contains an even number of quasiparticles\cite{Calabrese_2012} and occupies the Neveu-Schwarz sector.

Since we are interested in momentum space stabilizer entropies of the system, we apply an inverse Jordan-Wigner transformation\cite{nersesyan} from momentum space fermions, $c_k$ to momentum space qubits, 
$\sigma_k$'s.
The transformation is
\begin{gather}
\sigma^+_k=c^+_k\exp\left(i\pi\sum_{p<k}c^+_pc_p\right),\\
\sigma^z_k=2c^+_kc_k-1, \hspace*{4mm} \sigma^-_k=\left(\sigma^+_k\right)^+,
\end{gather}
and $\sigma^0_k$ is the identity.
For the Jordan-Wigner string operator $\exp\left(i\pi\sum_{p<k}c^+_pc_p\right)$, we need to order the momenta. This we achieve by taking a sequence of opposite momentum pairs as $-p_1,p_1,-p_2,p_2,-p_3,p_3$ etc.
 with $p_i>0$, 
but the adjacent pairs have no
particular order. With this, we cover all allowed momenta and in the sum in the exponent of the Jordan-Wigner string operator, all the momenta are included before $p=k$ using the above ordering procedure.
 
Using this transformation, the momentum space fermionic Hamiltonian maps onto a collection of interacting two qubit XY models 
in a perpendicular magnetic field with $k$ dependent exchange interaction and
 magnetic field as
\begin{gather}  
H=J\sum_{k>0} (g-\cos(k))\left(\sigma^z_k+\sigma^z_{-k}\right)-\nonumber\\
-i2\sin(k)\left(\sigma^+_k\sigma^+_{-k}-\sigma^-_k\sigma^-_{-k}\right),
\label{hamkspin}
\end{gather}
and constant terms are neglected. This model conserves parity of $z$ components of the total momentum space qubit operator expressed as $\sum_k\sigma^z_k$, similar to the approach in  Ref. \cite{laumann}.
The corresponding wavefunction, which is the qubit version of Eq. \eqref{wf}, is written in qubit basis as 
\begin{gather}
|\Phi\rangle=\bigotimes_{k>0}|\Phi_k\rangle,\hspace*{2mm}|\Phi_k\rangle =u_k|0\rangle_{-k}|0\rangle_{k}-iv_k |1\rangle_{-k}|1\rangle_k.
\label{wfspin}
\end{gather}
This momentum space basis represents the natural (and diagonal) basis deep in the paramagnetic phase since for $g\rightarrow \infty$, Eq. \eqref{hamk} becomes indeed diagonal in $k$. 
Since the stabilizer R\'enyi-entropies are basis dependent\cite{tarabunga},
the computational basis is the paramagnetic one. We note that real space magic has been probed in distinct bases\cite{haug} as well.

\paragraph{Pauli spectrum.}
In order to calculate the momentum space stabilizer R\'enyi entropies\cite{leone2022}, we need the expectation values of the momentum space Pauli spin operators for the ground state wavefunction.
These define the Pauli spectrum\cite{Beverland_2020,turkeshimagic} of the ground state wavefunction, expressed in momentum space in Eq. \eqref{wfspin}, in terms 
of the momentum space Pauli string. This is
\begin{gather}
\textmd{spec}\left(|\Phi\rangle \right)=\left\{ \langle \Phi|\Sigma|\Phi\rangle, \hspace*{1mm} \Sigma\in \Sigma_N\right\},
\end{gather}
where $\Sigma_N = \left\{\Sigma_{k_1}\otimes \Sigma_{k_2}\otimes\dots \Sigma_{k_N}\right\}$ are Pauli strings with $\Sigma_k=\sigma^{0,x,y,z}_k$ for a given momentum and there are $4^N$ different
$\Sigma$'s.
The Pauli spectrum, i.e. the distribution of Pauli strings is defined as
\begin{gather}
P(x)=4^{-N}\sum_{\Sigma\in \Sigma_N}\delta(x-|\langle \Phi|\Sigma|\Phi\rangle|).
\end{gather}
Here we only focus on the absolute value of the expectation values of Pauli strings because the stabilizer entropies depend
only on these\cite{leone2022}.
Since the wavefunction, Eq. \eqref{wfspin} factorizes into independent $(-k,k)$ momentum channels, it suffices to determine only the expectation values of $\sigma^\alpha_{-k}\sigma^\beta_k$ with {$\alpha,\beta\in\{0,x,y,z\}$.
The only non-vanishing expectation values are
\begin{subequations}
\begin{gather}
\langle\Phi_k|\sigma^0_{k}\sigma^0_{-k}|\Phi_k\rangle=\langle\Phi_k|\sigma^z_{k}\sigma^z_{-k}|\Phi_k\rangle=1,\\
\langle\Phi_k|\sigma^0_{k}\sigma^z_{-k}|\Phi_k\rangle=\langle\Phi_k|\sigma^z_{k}\sigma^0_{-k}|\Phi_k\rangle=v_k^2-u_k^2,\\
\langle\Phi_k|\sigma^x_{k}\sigma^y_{-k}|\Phi_k\rangle=\langle\Phi_k|\sigma^y_{k}\sigma^x_{-k}|\Phi_k\rangle=2u_kv_k,
\end{gather}
\label{matrixelements}
\end{subequations}
and the local Hilbert space of the $k$ and $-k$ qubits is $4$ dimensional. Within this subspace with 16 possible Pauli strings\cite{Beverland_2020}, only the above 6 of them give non-vanishing expectation values.
This means that for the whole system, there are $6^{N/2}$ non-vanishing Pauli string expectation values, and out of these, $2^{N/2}$ are unity, the remaining $4^N-6^{N/2}$ expectation values are zero.
The paramagnetic state with $\theta_k=0$ is a stabilizer state since the ferromagnetic state cannot be factorized in this basis. 
We note that for real space Pauli strings, the stabilizer states would be the real space 
ferromagnetic and ferromagnetic states. This hints that the two, real and momentum space approaches reveal a distinct but complementary face of
the underlying quantum model.

\begin{figure}[h!]
\centering
\includegraphics[width=7cm]{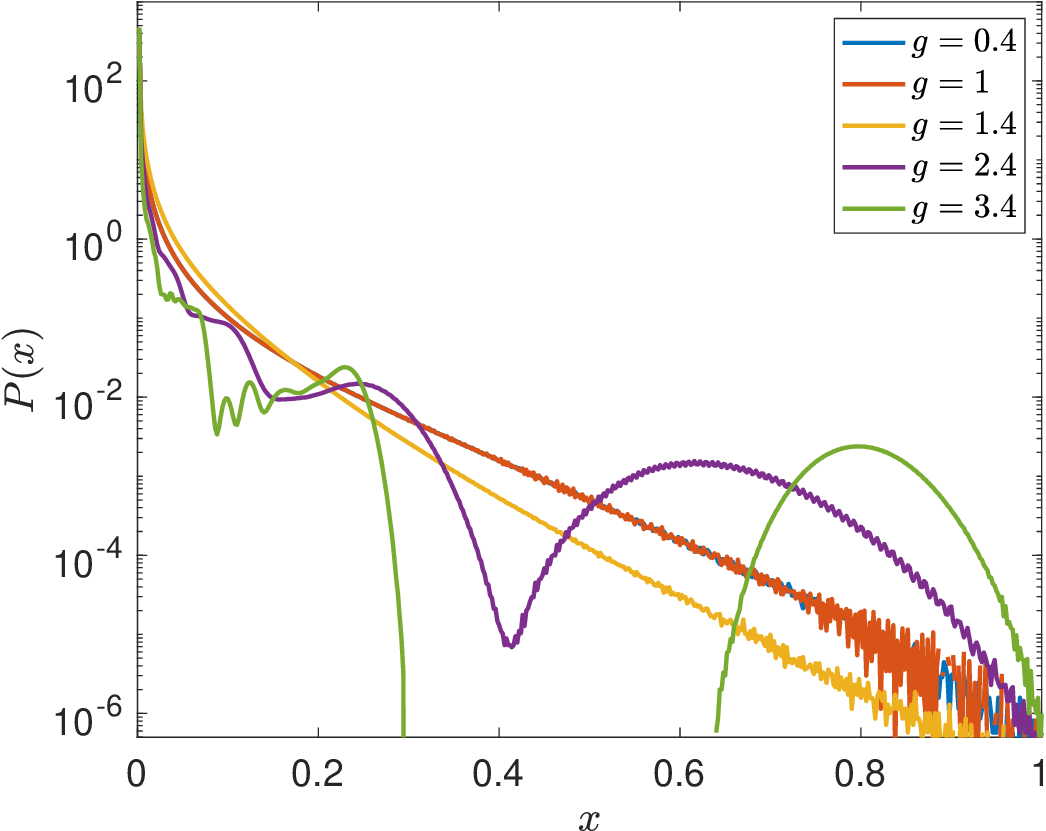}
\caption{The regular part of the distribution of non-zero Pauli strings, $|\langle\Phi|\Sigma|\Phi\rangle|\neq 0$ is plotted (normalized to unity here)
across the quantum phase transition for $N=40$. The $g=0.4$ and $g=1$ data fall on top of each other.
Since there are 3 different matrix elements for a $(k,-k)$ pair from Eq. \eqref{matrixelements}, effectively $3^{N/2}$ contributions were taken into account.
The Dirac delta peak at the origin from the vanishing strings, whose weight would be $1-(3/8)^{N/2}$, is not shown
for clarity. The overall weight of the non-zero strings would be $(3/8)^{N/2}$, and out of these, the weight of Pauli string expectation values of 1 is $8^{-N/2}$ from Eq. \eqref{matrixelements}.}
\label{fig:pauli}
\end{figure}

Using  Eq. \eqref{matrixelements} for the Pauli spectrum in a given $k$ and $-k$ channel, we evaluate the distribution of the Pauli spectrum of the whole system numerically, 
which is  shown in Fig. \ref{fig:pauli}. 

In the ferromagnetic phase, $g \le 1$, the distribution features a sharp peak around zero with an 
exponential tail that decays as $\ln(P(x)) \sim -N^{4/3}x$, and this behavior is independent of the transverse field amplitude $g$.
With increasing $g$  in the paramagnetic phase, the spectral weight redistributes, causing the peak around zero to sharpen and 
an additional structure (resonance) to emerge near $x=1$. This feature at $x=1$ slowly transforms into a delta peak as $g$ approaches 
$\infty$, leading to the vanishing of the stabilized entropies in the strong paramagnetic limit.
%
%
This is also understood from Eq. \eqref{matrixelements}: for $g\rightarrow\infty$, $\theta_k\approx\sin(k)/g\rightarrow 0$, 
therefore the three non-vanishing matrix elements for any momentum take the values
1, $-1$ and 0, respectively. Therefore, these are also the possible values what $\Sigma$ can take.

\paragraph{Magic gap.}
We also evaluate the magic gap\cite{bupnas,tarabungaprl}, which is the smallest distance of the available Pauli strings from one
as
\begin{gather}
\textmd{MG}\left(|\Phi\rangle \right)=1-\max_{\Sigma\in \Sigma_N, |\langle\Phi|\Sigma|\Phi\rangle|\neq 1} |\langle\Phi|\Sigma|\Phi\rangle|,
\end{gather}
which turns out to be zero. Even though the quasiparticle excitation spectrum is gapped away from the critical point and gapless exactly at the critical point,
the Pauli spectrum is gapless and the magic gap vanishes.
This is evident from the distribution close to one in Fig. \ref{fig:pauli}, and also follows from the explicit expression for the individual matrix elements from Eq. \eqref{matrixelements}.
Out of these, we need to find the largest below one. For the small momentum modes, $\theta_k\approx k/g$, therefore $(u_k,v_k)\rightarrow (1,0)$, thus the local Pauli string
 $|v_k^2-u_k^2|\rightarrow 1$.
By choosing all other local Pauli strings to be 1, we can approach 1 arbitrarily close with the total Pauli string, corroborating the gaplessness of the Pauli spectrum.

\paragraph{Stabilizer entropies.}
The stabilizer R\'enyi-entropies are associated to the probability distribution $|\langle \Phi|\Sigma|\Phi\rangle|^2/2^N$ with $\Sigma\in\Sigma_N$.
Using Eq. \eqref{matrixelements} and the structure of the wavefunction in Eq. \eqref{wf}, the stabilizer $n$-R\'enyi-entropies are given by
\begin{gather}
M_n=\frac{1}{1-n}\sum_{k>0}\ln\left(\frac{1+|\cos(\theta_k)|^{2n}+|\sin(\theta_k)|^{2n}}{2^{2n-1}}\right)-\nonumber\\
-N\ln(2).
\end{gather}
These entropies are constructed in such a way that they vanish for the stabilizer state\cite{leone2022} due to the last term.
Due to the momentum sum, they naturally follow a volume law. These are plotted in Fig. \ref{fig:stabrenyi} in the thermodynamic limit. Their values are constant within the ferromagnetic phase,
which follows from the magnetic field independence of the distribution of Pauli strings in this phase as shown in Fig. \ref{fig:pauli}.
At the critical point, they are non-analytic, develop a broad peak upon entering into the paramagnetic phase
and vanish with increasing transverse field.
\begin{figure}[t!]
\centering
\includegraphics[width=7cm]{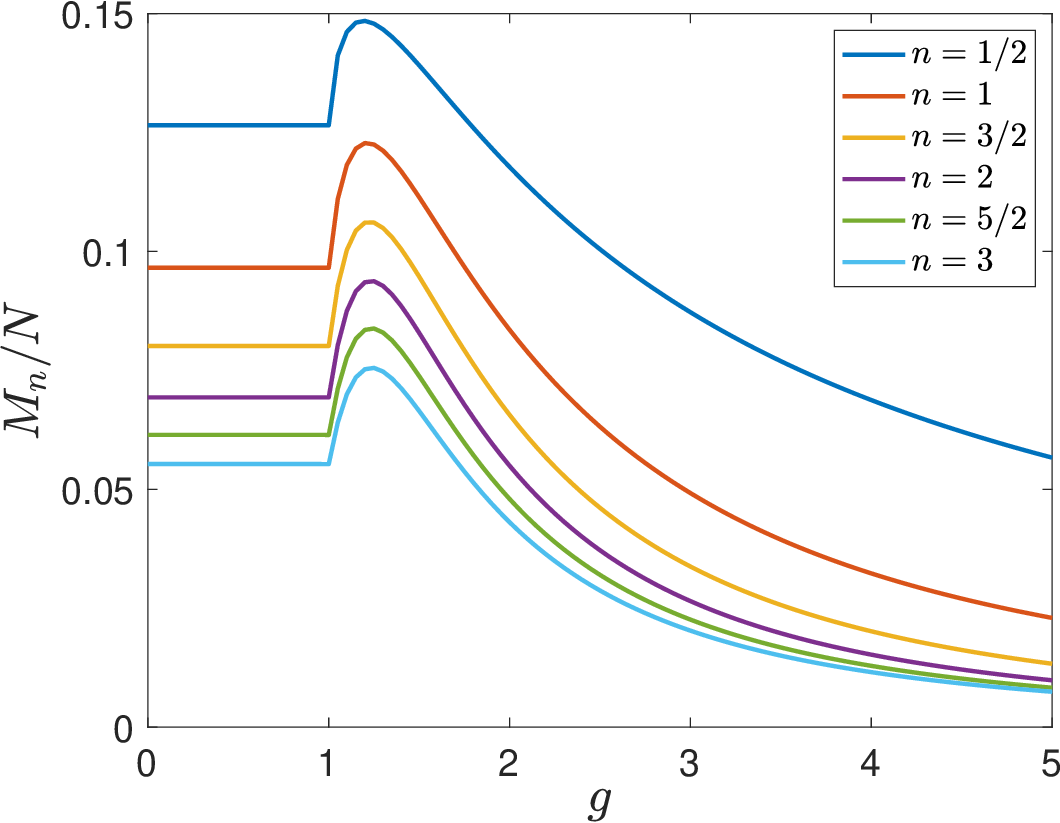}
\caption{The stabilizer R\'enyi entropies are plotted for $N=2000$ and several R\'enyi index $n$, displaying completely flat behaviour within the ferromagnetic phase, a broad 
peak upon entering into the paramagnetic phase
and vanishing in the $g\rightarrow\infty$ limit.}
\label{fig:stabrenyi}
\end{figure}

The $n=2$ stabilizer entropy a.k.a. magic is given by
\begin{gather}
M_2=-\sum_{k>0}\ln\left(\frac{7+\cos(4\theta_k)}{8}\right).
\end{gather}
Its value within the ferromagnetic phase in the thermodynamic limit is constant as 
\begin{gather}
\frac{M_2}{N}=-\frac{1}{2}\ln\left(\frac{7}{16} + \frac{\sqrt 3}{4}\right).
\end{gather}
The magic for various system sizes is plotted in Fig. \ref{fig:finitesize}. Finite size effect are sizeable in the ferromagnetic phase, 
which is the furthest away from the stabilizer state. Fluctuations are suppressed with increasing system size except for
the close vicinity of the $g=1$ critical point in the ferromagnetic side, where the thermodynamic limit is approached very slowly.
In the paramagnetic limit, the magic vanishes as $\sim 1/g^2$.
\begin{figure}[h!]
\centering
\includegraphics[width=7cm]{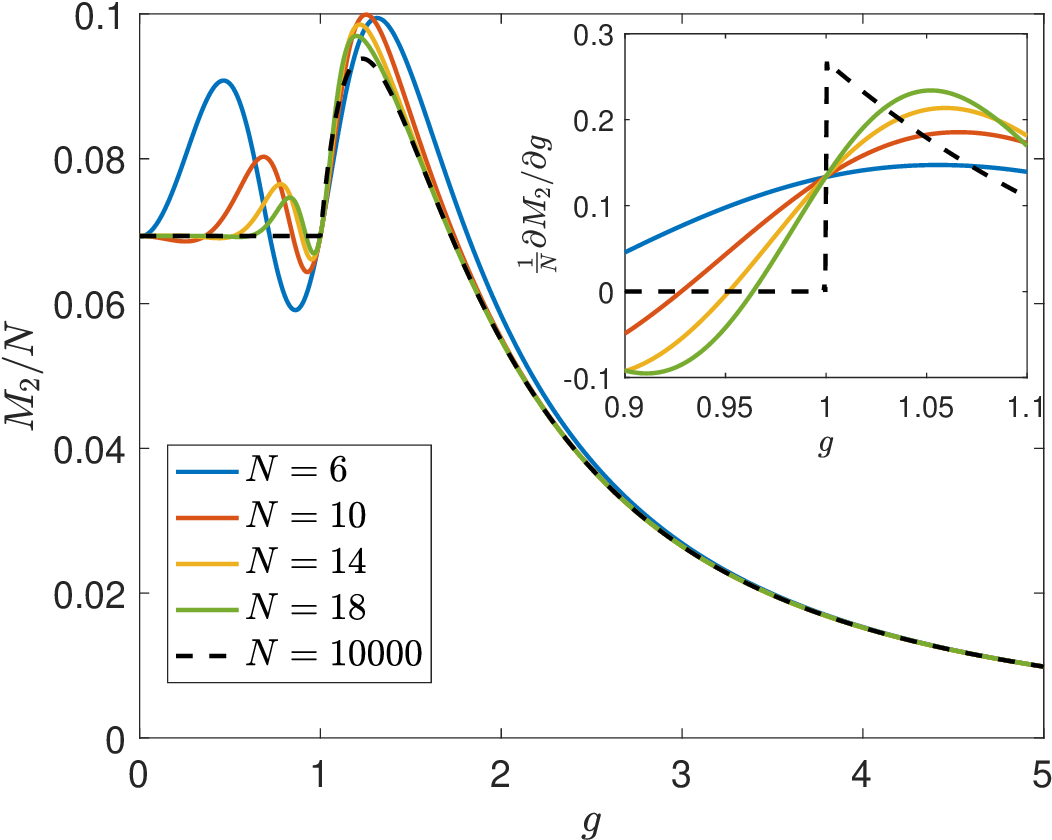}
\caption{The finite size scaling of the magic is shown for various system sizes. The inset zooms in to the derivative of the magic with respect to the transverse field at the quantum critical point.}
\label{fig:finitesize}
\end{figure}

The behaviour of these stabilizer entropies is in accord with the behaviour of the \emph{real space} magic\cite{olivieropra,tarabungaprl}. Similar behaviour is expected on general ground from thermodynamic entropies at around critical points. 
They are non-analytic function at $g=1$, and their first derivative
with respect to the transverse field jumps at $g=1$, indicating a continuous quantum phase transition. The size of the jump is $\frac 1N \lim_{g\rightarrow 1^+}\partial M_2/\partial g=2-\sqrt 3$ in the thermodynamic limit,
and $\lim_{g\rightarrow 1^-}\partial M_2/\partial g=0$ as evident from Fig. \ref{fig:finitesize}.
This is analogous to the slope change of the thermodynamic entropy at a second order thermal phase transition and the concomitant jump in the specific heat at the critical temperature\cite{herbutbook}.

With this knowledge, we  compare real space and momentum space magic and nonstabilizerness. The real space magic per site at the critical point, where quantum fluctuations are the most enhanced, 
is\cite{leone2022,lami} $\approx 0.31$, while
the corresponding momentum space magic is much smaller,  $\approx 0.07$. This is further reduced in the ferromagnetic phase of  finite size systems, see Fig. \ref{fig:finitesize}.
This indicates that the transverse field Ising chain can be more efficiently simulated classically in momentum space due 
to its reduced  magic. However, away from the quantum critical point, the momentum space advantage melts away rapidly, especially in the absence of transverse field,
when the momentum space magic is still finite but the real space magic vanishes. 

\paragraph{Discussions.}

In terms of future investigations, it would be interesting to extend this study to finite temperatures as well as considering departures from exact solvability.
 Dynamical properties of the system and the occurrence of dynamical quantum phase transition\cite{heyl} in momentum space magic would be another promising 
line of research, together with  momentum space magic of higher dimensional models\cite{kitaev2006}.  

In summary, we studied the momentum space nonstabilizerness of the transverse field Ising chain. By mapping it to momentum space qubits, we find that the distribution of 
the absolute value of the Pauli strings displays a broad distribution in the ferromagnetic phase, independent from the transverse field, which evolves 
into a two peaked distribution (around 0 and 1) in the paramagnetic phase. The magic gap vanishes for all values of the transverse field and the Pauli spectrum is gapless.
In the thermodynamic limit, all stabilizer entropies are independent from the transverse field in the ferromagnetic phase, indicating that all these states are equally magical.
They display non-analytic behaviour at the critical point and 
vanish deep in the paramagnetic phase upon approaching the stabilizer state. The momentum space magic at the critical point is much smaller compared to its real space counterpart,
therefore classical simulations  in momentum space are more efficient.

\begin{acknowledgments}
Discussions with M. Kormos and G. Tak\'acs are gratefully acknowledged.
This research is supported by the National Research, Development and 
Innovation Office - NKFIH  within the Quantum Technology National Excellence 
Program (Project No.~2017-1.2.1-NKP-2017-00001), K134437, K142179 by the BME-Nanotechnology 
FIKP grant (BME FIKP-NAT), the QuantERA ‘QuSiED’ grant No. 101017733, 
and by a grant of the Ministry of Research, Innovation and
 Digitization, CNCS/CCCDI-UEFISCDI, under projects number PN-IV-P1PCE-2023-0159 and PN-IV-P1PCE-2023-0987.
\end{acknowledgments}

\bibliographystyle{apsrev}
\bibliography{wboson1}

\end{document}